\documentclass[12pt]{article}
\usepackage{amsmath}
\usepackage{amssymb}
\usepackage{amsfonts}
\usepackage{graphicx}
\newcommand{\be}{\begin{equation}}
\newcommand{\ee}{\end{equation}}
\newcommand{\bea}{\begin{eqnarray}}
\newcommand{\eea}{\end{eqnarray}}

\begin{document}

\bigskip\begin{titlepage}
\begin{flushright}
UUITP-29/08\\
NORDITA-2008-67\\
\end{flushright}
\vspace{1cm}
\begin{center}
{\Large\bf Black holes in asymptotically Lifshitz spacetime\\}
\end{center}
\vspace{3mm}
\begin{center}
{\large
Ulf H.\ Danielsson$^{1,a}$ and L\'arus Thorlacius$^{2,3,b}$} \\
\vspace{5mm}
1) Institutionen f\"or fysik och astronomi \\
Uppsala Universitet,
Box 803, SE-751 08
Uppsala, Sweden \\
\vspace{5mm}
2) NORDITA \\
Roslagstullsbacken 23, SE-106 91 Stockholm, Sweden \\
\vspace{5mm}
3) Science Institute \\
University of Iceland, Dunhaga 3, IS-107 Reykjavik, Iceland \\
\vspace{5mm}
{\tt
{$^a\>$}ulf.danielsson@physics.uu.se \qquad
{$^b\>$}larus@nordita.org \\
}
\end{center}
\vspace{5mm}
\begin{center}
{\large \bf Abstract}
\end{center}
\noindent
A model of 3+1 dimensional gravity with negative cosmological constant coupled to abelian
gauge fields has been proposed as a gravity dual for Lifshitz like critical phenomena in
2+1 dimensions. The finite temperature behavior is described by black holes that are 
asymptotic to the Lifshitz fixed point geometry. There is a one-parameter family of charged
black holes, where the magnitude of the charge is uniquely determined by the black hole area. 
These black holes are thermodynamically stable and become extremal in the limit of vanishing 
size. The theory also has a discrete spectrum of localized objects described by non-singular 
spacetime geometries. The finite temperature behavior of Wilson loops is reminiscent of strongly coupled gauge theories in 3+1 dimensions, including screening at large distances.
\vfill
\begin{flushleft}
December 2008
\end{flushleft}
\end{titlepage}\newpage

\section{Introduction}

AdS/CFT duality 
\cite{Maldacena:1997re,Gubser:1998bc,Witten:1998qj} provides unique access
to the physics of supersymmetric gauge theories at strong coupling. 
The original conjecture, and various refinements of it, have been tested to the 
extent that its validity is generally accepted in the theoretical high-energy 
physics community. By now AdS/CFT duality has been generalized in many
directions and the notion of a gauge theory/gravity correspondence has become 
a standard item in the toolbox of high-energy theory. Recently, considerable
effort has been put into extending AdS/CFT beyond high-energy physics by 
constructing gravity models that are conjectured to be dual to various condensed 
matter systems \cite{Herzog:2007ij}-\cite{Basu:2008st}. In this case there is
no supersymmetry, or truncation of higher-dimensional supergravity, to justify the
duality but this is not necessarily a problem. The gravity dual provides a 
phenomenological description of whatever strongly coupled physics that is 
being modeled, and, as such, it can be useful even if the connection to the 
underlying dynamics cannot be spelled out in detail. Ultimately, the usefulness 
of a dual description of a real physical system is to be judged by its success in 
explaining, or better yet, predicting experimental results.

In the present paper we develop further a recently proposed gravitational dual 
description \cite{Kachru:2008yh} of a class of critical phenomena exhibiting
unconventional scaling of the form
\be
t\rightarrow\lambda^{z}t,\qquad\mathbf{x}\rightarrow\lambda\mathbf{x},
\label{lscaling}
\ee
with $z\neq1$. The so called Lifshitz theory,
\be
L=\int d^{2}xdt\left(  \left(  \partial_{t}\phi\right)  ^{2}-K\left(
\nabla^{2}\phi\right)  ^{2}\right)  .
\label{laction}
\ee
provides a simple example of a 2+1 dimensional field theory which is invariant 
under precisely this kind of scaling (with $z=2$). This, and other related models, 
have been used to model quantum critical behavior in strongly correlated 
electron systems \cite{Rokhsar-1988}-\cite{vishwanath-2003}.

In \cite{Kachru:2008yh} it was conjectured that strongly 
coupled systems with Lifshitz scaling can be modeled by a gravity theory 
with a spacetime metric of the following form
\begin{equation}
ds^{2}=L^{2}\left(  -r^{2z}dt^{2}+r^{2}d^{2}\mathbf{x}+\frac{dr^{2}}{r^{2}
}\right)  .
\label{lmetric}
\end{equation}
This metric is invariant under the transformation
\begin{equation}
t\rightarrow\lambda^{z}t,\quad r\rightarrow\frac{r}{\lambda},\quad
\mathbf{x}\rightarrow\lambda\mathbf{x}\,.
\end{equation}
The coordinates $(t,r,x^{1},x^{2})$ are dimensionless and the only
characteristic length scale of the geometry is $L$. It was shown in
\cite{Kachru:2008yh} that an action coupling four-dimensional gravity, with a
negative cosmological constant, to a simple complement of abelian gauge
fields,
\begin{equation}
S=\int d^{4}x\sqrt{-g}\left(  R-2\Lambda\right)  -\frac{1}{2}\int\left(
F_{\left(  2\right)  }\wedge\ast F_{\left(  2\right)  }+H_{\left(  3\right)
}\wedge\ast H_{\left(  3\right)  }\right)  -c\int B_{\left(  2\right)  }\wedge
F_{\left(  2\right)  },\label{action}
\end{equation}
can support such a metric with $z>1$. Here $F_{\left(  2\right)  }=dA_{\left(
1\right)  }$ and $H_{\left(  3\right)  }=dB_{\left(  2\right)  }$ are a
two-form and a three-form field strength respectively, and the length scale
$L$ is related to the cosmological constant $\Lambda=-5/L^{2}$. 

Here we focus on the $z=2$ case but we expect that most of our results
can be generalized to other $z$ values. In section 2, we extend the
analysis of \cite{Kachru:2008yh} to global coordinates, and demonstrate the
existence of a discrete set of solutions that we call Lifshitz stars. We
also introduce finite temperature by having a black hole at the center of an
asymptotically Lifshitz space time. Such a black hole carries an electric charge
that couples to the two-form gauge field strength. Interestingly, the magnitude
of the charge is uniquely fixed by the black hole size and these black holes
become extremal in the small size limit. In section 3 we study the
thermodynamic properties of black holes in asymptotically Lifshitz spacetime, 
and conclude that, contrary to black holes in asymptotically AdS spacetime, 
they are thermodynamically stable for all black hole sizes. In section 4 we 
speculate on the nature of the boundary theory, and perform a calculation
analogous to the evaluation of a Wilson loop in AdS/CFT. 

A number of recent papers have analyzed black hole geometries in 
gravity duals of non-relativistic quantum systems 
\cite{Herzog:2008wg}-\cite{Yamada:2008if} but those gravitational models
are different from the one we consider here, leading to a different spectrum
of black holes and different thermodynamic properties.

\section{Asymptotically Lifshitz spacetime}

In this section we look for global metrics that solve the equations of motion of (\ref{action})
and approach the Lifshitz geometry (\ref{lmetric}) in an asymptotic limit. We find two types 
of spherically symmetric, static solutions. One is a black hole with a non-degenerate event 
horizon and the other is a smooth geometry that describes a 
non-singular, spherically symmetric concentration of the gauge fields. 

The equations of motion of (\ref{action}) are easily obtained. The gauge
fields satisfy a pair of coupled equations,
\begin{align}
d\ast F_{\left(  2\right)  } &  =-cH_{\left(  3\right)  },\label{Feq}\\
d\ast H_{\left(  3\right)  } &  =cF_{\left(  2\right)  },\label{Heq}
\end{align}
and the Einstein equations are
\begin{equation}
G_{\mu\nu}-\frac{5}{L^{2}}g_{\mu\nu}=\frac{1}{2}(F_{\mu\lambda}F_{\nu
}^{\>\lambda}-\frac{1}{4}g_{\mu\nu}F_{\lambda\sigma}F^{\lambda\sigma}
)+\frac{1}{4}(H_{\mu\lambda\sigma}H_{\nu}^{\>\lambda\sigma}-\frac{1}{6}
g_{\mu\nu}H_{\lambda\sigma\rho}H^{\>\lambda\sigma\rho}).
\end{equation}
The Lifshitz metric (\ref{lmetric}) is a solution of the equations of motion
if  the topological coupling between the gauge fields is tuned to be
\begin{equation}
c=\frac{\sqrt{2z}}{L}.\label{cvalue}
\end{equation}
We will assume this value for $c$ in what follows.  

Our ansatz for the metric 
and the two- and three-form fluxes generalizes the one employed
in \cite{Kachru:2008yh} to global coordinates,
\begin{eqnarray}
ds^{2}&=&L^{2}\left(  -r^{4}f\left(  r\right)  ^{2}dt^{2}+r^{2}d^{2}\Omega
+\frac{g\left(  r\right)  ^{2}}{r^{2}}dr^{2}\right)  ,
\label{almetric} \\
F_{\left(  2\right)  }  &=& \frac{2}{L}h(r)\, \theta_{r}\wedge\theta_{t} , \\
H_{\left(  3\right)  }  &=& \frac{2}{L}j(r)\, \theta_{r}\wedge\theta_{\theta}
\wedge\theta_{\phi},
\end{eqnarray}
where
\begin{eqnarray}
\theta_{r}  &=& L\frac{g\left(  r\right)  }{r}dr , \\
\theta_{t}  &=& Lr^{2}f\left(  r\right)  dt , \\
\theta_{\theta}  &=& Lrd\theta , \\
\theta_{\phi}  &=& Lr\sin\theta d\phi .
\end{eqnarray}
The two-form field strength $F_{\left(  2\right)  }$, given by $h\left(  r\right)  $, is an 
electric field directed radially outwards from the origin at $r=0$. The topological coupling 
between $F_{(2)}$ and $H_{(3)}$ in the action implies that the three-form flux is 
electrically charged and acts as a source of the electric field. The above ansatz
for $H_{\left(  3\right) }$ thus corresponds to a charged fluid whose density is governed
by $j\left(  r\right)$.

The Einstein equations and the field equations for the gauge fields 
reduce to a system of non-linear first order differential equations,
\begin{eqnarray}
r f^{\prime}  &=& -\frac{5f}{2} + \frac{fg^{2}}{2}\left(  5+\frac{1}{r^2}+j^{2}-h^{2}\right) , \label{firsteq} \\
rg^{\prime}  &=& \frac{3g}{2}-\frac{g^{3}}{2}\left(5+\frac{1}{r^2} - j^{2}-h^{2}\right)  , \label{secondeq} \\
rj^{\prime}  &=& 2gh+\frac{j}{2}-\frac{jg^{2}}{2}\left(5+\frac{1}{r^2} +j^{2}-h^{2}\right), \label{thirdeq}  \\
rh^{\prime}  &=& 2gj-2h \label{fourtheq} .
\end{eqnarray}
In the $r\gg 1$ limit, we can replace the sphere metric, $d^{2}\Omega,$ in (\ref{almetric}) by a 
flat metric, or, equivalently, neglect the $1/r^2$ terms in equations (\ref{firsteq}) to (\ref{thirdeq}). 
In this case we recover the Lifshitz solution, 
$f\left(  r\right) = g\left(  r\right)  =h\left(  r\right)  =j\left(  r\right)  =1$, 
as constructed by \cite{Kachru:2008yh}. 
It is also easy to see that the AdS-Schwarzschild geometry,
\be
g(r) = \frac{r}{\sqrt{\frac{5}{3}r^2+1-\frac{\mu}{r}}}, \qquad 
f(r) = \frac{\sqrt{\frac{5}{3}r^2+1-\frac{\mu}{r}}}{r^2} ,
\ee
is a solution of the equations
with $h(r)=j(r)=0$ but we have not found explicit analytic solutions with non-trivial gauge fields. 
It is, however, straightforward to integrate the system of equations numerically and one can 
learn a lot about the behavior of solutions by a combination of numerics and asymptotic 
analysis. 

It is convenient to first consider equations (\ref{secondeq}) - (\ref{fourtheq}), which only involve
the three functions $g(r)$, $h(r)$, and $j(r)$ and then, given a solution to this system, solve 
equation (\ref{firsteq}) for the remaining function $f(r)$. 
We are primarily interested in geometries that are asymptotically Lifshitz in the sense that
$f,g,h,j\rightarrow 1$ as $r\rightarrow\infty$. In order to study the asymptotic behavior at 
large $r$ we linearize the system (\ref{secondeq}) - (\ref{fourtheq}) around the fixed point,
\be
r \frac{d}{dr}\left[
\begin{array}{c}
\delta g \\
\delta h \\
\delta j
\end{array}
\right]=\left(
\begin{array}{rrr}
-3 & 1 & 1 \\
2 & -2 & 2 \\
-3 & 3 & -3
\end{array}
\right) \left[
\begin{array}{c}
\delta g \\
\delta h \\
\delta j
\end{array}\right]-\frac{1}{2r^2}
\left[
\begin{array}{c}
1 \\
0 \\
1
\end{array}\right],
\label{lineareqs}
\ee
where $g=1+\delta g$, {\it etc.} 
By standard manipulations, the matrix that appears in the linear system 
can be brought into Jordan form,
\be
\left(
\begin{array}{rrr}
-4 & 1 & 0 \\
0 & -4 & 0 \\
0 & 0 & 0
\end{array}
\right),
\ee
from which we read off that at the linear level there are two decaying modes,
that behave as $1/r^4$ and $\log(r)/r^4$, and a zero mode that does
not depend on $r$. In addition to the eigenmodes, the general solution of
the linear system includes a universal $1/r^2$ mode that comes from the 
inhomogeneous term on the right 
hand side of (\ref{lineareqs}). It is straightforward to find the general solution 
of the linear system and confirm that it has all these features, but, since we
do not need the details of the full solution for what follows, we do not write
it down here.

At large values of $r$ the non-zero modes, including the universal 
 $1/r^2$ mode, have decayed away leaving only the zero mode behind.
 The solution then has the form
 \be
 g\approx 1+\gamma, \qquad h\approx 1+2\gamma, 
\qquad j\approx 1+\gamma,
\label{zeromode}
\ee
with $\gamma\ll 1$,
but this is not the whole story. When non-linear corrections are included,
the zero mode is lifted and becomes either a marginally growing or marginally
decaying mode. 

To study the evolution of the zero mode at large $r$, we insert (\ref{zeromode}) into the 
original non-linear  equations of motion (\ref{secondeq}) - (\ref{fourtheq}), except that, 
since we are assuming that the non-zero modes have already decayed, we drop the 
$1/r^2$ terms in the equations. Working to leading non-vanishing order in $\gamma$ 
then gives
\be
r  g'(r)\approx 7\gamma^2, \qquad r h'(r)\approx 2\gamma^2, 
\qquad r j'(r)\approx \gamma^2.
\label{zeromodeeqs}
\ee
The leading order terms on the right hand side all come with a positive sign. This immediately
implies that a zero mode of positive amplitude $\gamma>0$ is a growing mode and a
solution of the equations of motion where such a mode is present at large $r$ cannot be 
asymptotic to the Lifshitz fixed point.

A zero mode of negative amplitude will, on the other hand, slowly decay. The decay can
be described analytically in a simple fashion. Writing $\gamma(r)=-1/\log{r}$ it is
straightforward to check that 
\begin{eqnarray}
g(r) &\approx& 1+\gamma(r)+\gamma^2(r)+\ldots ,  \nonumber \\
h(r) &\approx& 1+2\gamma(r)-\gamma^2(r)+\ldots ,  \\
j(r) &\approx& 1+\gamma(r)-2\gamma^2(r)+\ldots ,  \nonumber 
\end{eqnarray}
solves the non-linear equations of motion up to terms that are small compared to 
$1/(\log r)^2$. The solution thus appears to be asymptotic to the Lifshitz 
fixed point, at least as far as the functions $g$, $h$, and $j$ go, but the approach 
to the fixed point is extremely slow. In order to decide whether the geometry is
truly asymptotic to the Lifshitz geometry we have to consider the evolution of 
the remaining function $f(r)$, which enters into the $g_{tt}$ component of the
metric (\ref{almetric}). Inserting a zero mode (\ref{zeromode}) with 
$\gamma(r)=-1/\log{r}$ into the $f(r)$ equation of motion (\ref{firsteq}) gives
\be
r f'(r) = -\frac{4f(r)}{\log r}+\ldots .
\ee
This integrates to
\be
f(r) = \frac{\textrm{const}}{(\log r)^4},
\ee
which goes to zero in the $r\rightarrow\infty$ limit. A solution with a marginally decaying 
zero mode is therefore not asympotic to the Lifshitz fixed point after all.

The upshot of all this is that the only solutions of our system that are asymptotic to the 
Lifshitz fixed point are those for which the amplitude of the zero mode of the
linearized system happens to vanish in the asymptotic region. This requires fine-tuning
of initial values when solving the non-linear equations of motion, which in turn reduces
the number of free parameters in the solutions that are of most interest to us.

\subsection{Black holes}

\begin{figure}
\begin{center}
\includegraphics[
width=3.5in
]
{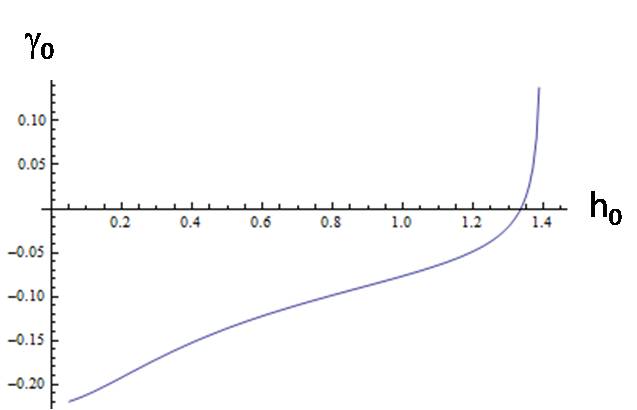}
\end{center}
\caption{\small Large $r$ behavior of numerical black hole solutions.
The figure plots the amplitude of the zero mode $\gamma_0$ at $r=10^6$ as a function of 
$h(r_0)$, the electric field at the horizon, for a black hole with $r_0=10$. The curve has a 
single zero at which the zero mode amplitude vanishes, and this uniquely determines a
value of $h(r_0)$ for which the black hole is asymptotic to the Lifshitz fixed point geometry.}
\label{zeromodeplot}
\end{figure}

We now describe results obtained by integrating the equations of motion numerically.
We find a one-parameter family of black hole solutions with a non-degenerate horizon
which are asymptotic to the Lifshitz geometry (\ref{lmetric}). The characteristic parameter 
of the black hole can be either taken as $r_0$, the value of the area coordinate $r$ at
the horizon, or $h_0 \equiv h(r_0)$, the value of the radial electric field at the horizon.
At first sight, one would expect these two parameters to be independent as they are
for an ordinary Reissner-Nordstr\"om black hole but, as discussed above, only a 
restricted set of geometries approaches the Lifshitz fixed point at $r\rightarrow\infty$.
The restriction on the parameters can be understood in terms of the interaction 
between the charged fluid, represented by the three-form field strength, and the
black hole. If a neutral black hole is placed within a charged fluid, then some 
of the fluid flows into the black hole and makes it charged. A static geometry 
describes an equilibrium configuration of the charged fluid outside a charged black 
hole, where the electric repulsion from the charge of the black hole precisely balances 
the gravitational pull.

Let us assume that there is a non-degenerate horizon at $r=r_0$. The $g_{tt}$ component 
of the metric should then have a simple zero and the $g_{rr}$ component a simple pole at 
the horzion. If we further assume that the electric field $h(r)$ has a finite value at the 
horizon, we find that the charged fluid density must go to zero at the horizon. This is 
in line with the equilibrium argument in the previous paragraph.

With these assumptions, we can develop a near-horizon expansion of the various fields,
\begin{eqnarray}
f(r) &=&\sqrt{r-r_0}\left( f_0+f_1(r-r_0)+f_2(r-r_0)^2+\ldots \right), \nonumber \\ 
g(r) &=& \frac{1}{\sqrt{r-r_0}}\left(g_0+g_1(r-r_0)+g_2(r-r_0)^2+\ldots \right) , \nonumber \\
j(r) &=&\sqrt{r-r_0}\left( j_0+j_1(r-r_0)+j_2(r-r_0)^2+\ldots \right), \nonumber \\
h(r) &=& h_0+h_1(r-r_0)+h_2(r-r_0)^{2}+\ldots .
\label{expansion}
\end{eqnarray}
Inserting this into the equations of motion and working order by order in $r-r_0$ one
obtains relations between the various constant coefficients,
\begin{eqnarray}
g_0 &=& \frac{r_0^{3/2}}{\sqrt{(5-h_0^2)r_0^2+1}}, \\  
j_0 &=& \frac{2h_0\sqrt{r_0}}{\sqrt{(5-h_0^2)r_0^2+1}}, \\
h_1 &=& \frac{2h_0r_0((h_0^2-3)r_0^2-1)}{(5-h_0^2)r_0^2+1}, \\
f_1 &=& \frac{f_0((6h_0^4-52h_0^2+100)r_0^4+(45-11h_0^2)r_0^2+5)}{2r_0((5-h_0^2)r_0^2+1)^2}, \\
\  &\vdots & \ \nonumber 
\end{eqnarray}
We notice that for a given black hole size, $r_0$, there is an upper bound on the 
electric field strength at the horizon,
\be
\vert h_0\vert < \sqrt{5+\frac{1}{r_0^2}} ,
\label{upperbound}
\ee
beyond which $g_0$ and $j_0$ would be complex valued. 

\begin{figure}
\begin{center}
\includegraphics[
height=2.5in,
]
{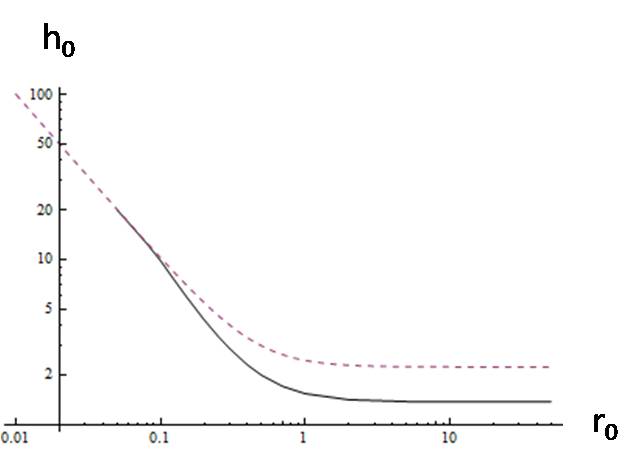}
\end{center}
\caption{\small The charge of an asymptotically Lifshitz black hole is uniquely determined
by the horizon area. The figure plots $h(r_0)$, the electric field strength at the horizon, 
as a function of $r_0$, the area coordinate at the horizon. The dashed curve is the upper
bound on $h_0$ in equation (\ref{upperbound}). The bound is saturated in the small black
hole limit.}
\label{hvsr0}
\end{figure}

We now use the expansion to generate initial values for the numerical integration of the 
equations of motion (\ref{firsteq}) - (\ref{fourtheq}), starting close to the horizon and 
integrating outwards in $r$. We have a two-parameter family of initial data, using $r_0$ and 
$h_0$ as the independent variables in the expansion. At first sight, $f_0$ also appears to
be an independent free parameter but this is not really the case. Equation (\ref{firsteq}) is 
linear in $f$ so the overall normalization of $f(r)$ is not determined. As discussed above, 
the solutions that are of most interest to us are those where $f$ goes to a constant 
asymptotically, $f(r)\rightarrow f_\infty$ as $r\rightarrow\infty$. Such a solution can always 
be normalized to $f\rightarrow 1$ by dividing through by $f_\infty$. We can therefore set 
$f_0=1$ in our numerical runs and take care of the normalization of $f$ at the end of the day.

The next step is to look systematically for solutions that describe black holes in an 
asymptotically Lifshitz background. A convenient way to conduct the search is to fix
$r_0$ and then prepare a sequence of initial data for different values of $h_0$. 
For each set of initial values, the equations of motion are integrated from the
near-horizon region out to sufficiently large $r$ so that the non-zero eigenmodes have
decayed away. One then looks for a zero mode of the form (\ref{zeromode}) and notes
how the amplitude $\gamma_0$ varies as a function of $h_0$, for a given value of $r_0$.
The result of this procedure is shown in Figure~\ref{zeromodeplot}. 
There is a unique value of $h_0$, for which the amplitude of the zero mode vanishes, 
and this corresponds to the charge of
an asymptotically Lifshitz black hole with horizon at $r=r_0$. 

This can all be repeated for different sized black holes and Figure~\ref{hvsr0} shows the 
critical value of $h_0$ as a function of $r_0$. This figure nicely summarizes the 
numerical results of this subsection. Starting from a two-parameter family of 
initial data, we have found a one-parameter family of black holes geometries that have
the correct asymptotic behavior to be sitting in a Lifshitz background. The upper bound
(\ref{upperbound}) on $h_0$ is given by the dashed curve in the figure. The bound is 
saturated, {\it i.e.} the black hole is becoming extremal, in the small black hole limit.

\begin{figure}
\begin{center}
\includegraphics[width=2.9in]{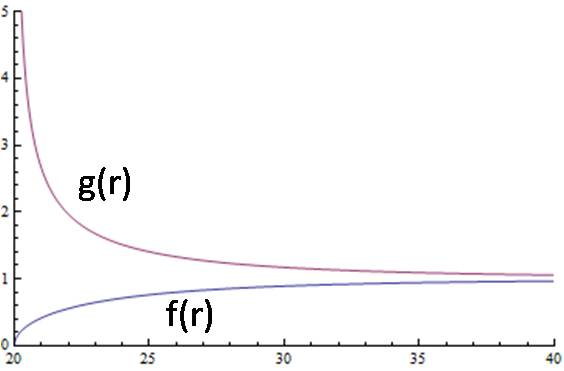}
\includegraphics[width=2.9in]{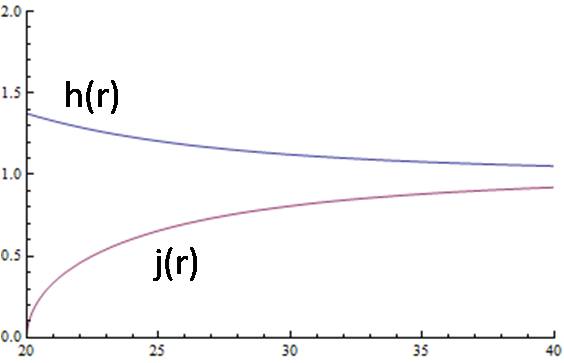}
\end{center}
\caption{\small A large black hole with $r_0=20$. The figure on the left shows the 
metric functions $f(r)$ and $g(r)$, while the figure on the right shows the electric 
field $f(r)$ and the charge density $j(r)$.}
\label{largebh}
\end{figure}

\begin{figure}
\begin{center}
\includegraphics[
width=2.9in
]
{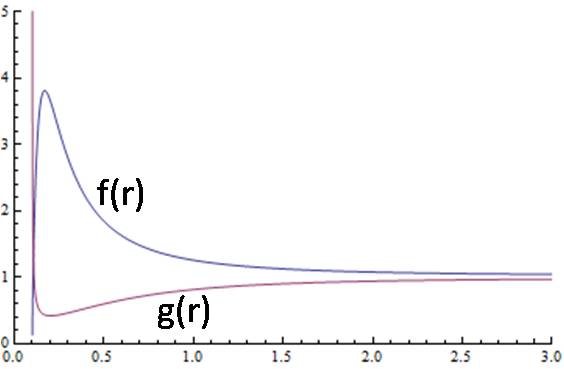}
\includegraphics[
width=2.9in
]
{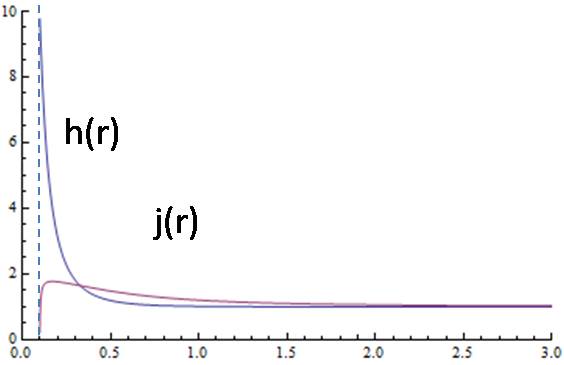}
\end{center}
\caption{\small A small black hole with $r_0=0.1$. The figure on the left shows the 
metric functions $f(r)$ and $g(r)$, while the figure on the right shows the electric 
field $f(r)$ and the charge density $j(r)$.}
\label{smallbh}
\end{figure}

The actual metric and gauge field configurations for two such black holes, a large one 
and a small one, are shown in Figures~\ref{largebh} and~\ref{smallbh} 
respectively.

\subsection{Lifshitz stars}\label{lstars}

\begin{figure}
\begin{center}
\includegraphics[
natheight=2.500200in,
natwidth=3.760200in,
height=2.5417in,
width=3.8078in
]
{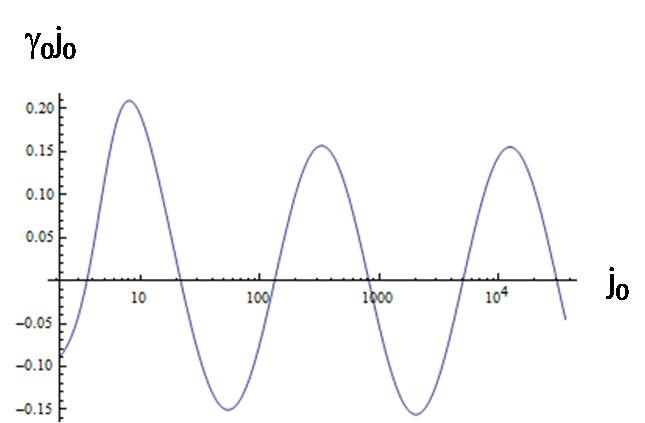}
\end{center}
\caption{\small Large $r$ behavior of the non-singular solutions discussed in Section~\ref{lstars}. 
The figure indicates the amplitude of the zero mode $\gamma_0$ at $r=10^4$ as a function of 
$j_0$, the charge density at the origin. The zeroes correspond to Lifshitz stars which are
asymptotic to the Lifshitz fixed point geometry. We plot the product $j_0\gamma_0$ rather than 
$\gamma_0$ itself. This does not affect the location of the zeroes but makes the plot more
readable. }
\label{starzero}
\end{figure}

In this subsection we turn our attention to a different kind of localized object in an
asymptotically Lifshitz background, one described by a smooth geometry with no horizon
or curvature singularity anywhere in spacetime. It has a spherically symmetric concentration
of the charged fluid with $j(0)$ taking a finite value. Since the fluid density is finite 
at the origin the electric field strength must vanish there, which translates into $h(0)= 0$.
If we also impose that the $g_{tt}$ and $g_{rr}$ components of the metric be finite
at $r=0$ we arrive at the following expansion at small $r$,
\begin{eqnarray}
j(r) &=& j_0\left(1 -\frac{1}{6}(1+2j_0^2) r^2+\frac{1}{360}(103+152j_0^2+36j_0^4) r^4+\ldots 
\right), \\
h(r) &=& j_0 r \left(\frac{2}{3}-\frac{1}{15}(6+j_0^2) r^2+\frac{1}{1260}(528+96j_0^2+7j_0^4) r^4
+\ldots \right), \\
g(r) &=& r \left(1 +\frac{1}{6}(-5+j_0^2) r^2+\frac{1}{360}(375-146j_0^2-9j_0^4) r^4+\ldots 
\right), \\
f(r) &=&\frac{1}{r^2}\left(1 +\frac{1}{6}(5+2j_0^2) r^2+\frac{1}{360}(-125+16j_0^2+4j_0^4) r^4+\ldots 
\right).
\end{eqnarray}
This can be used to generate a one-parameter family of initial value data, corresponding
to different charge densities $j_0$ at the origin. We start the numerical integration at  
small $r$ and integrate outwards to a large value of $r$, where the non-zero eigenmodes
have decayed away. Figure~\ref{starzero} displays the amplitude of the remaining zero mode as a
function of $j_0$. It reveals a discrete set of `magic' values of $j_0$ for which the 
zero-mode vanishes and the geometry is asymptotic to the Lifshitz fixed point geometry.

We refer to the configurations with magic $j_0$ values as Lifshitz stars. They occur at
$j_0=3.59,\>21.8,\>1.34\times 10^2,\>8.18\times 10^2,\>5.05\times 10^3,\>2.99\times 10^4,\ldots$. 
These magic values can be seen in Figure~\ref{starzero} and the shape of the curve in
the figure suggests that the sequence continues. We eventually run out of numerical precision 
when we go to higher $j_0$ values. 

\begin{figure}
\begin{center}
\includegraphics[
width=2.9in
]
{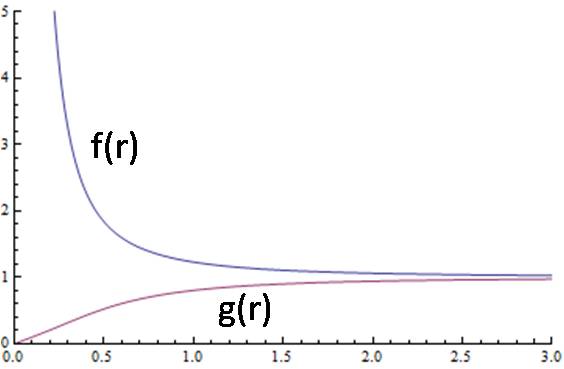}
\includegraphics[
width=2.9in
]
{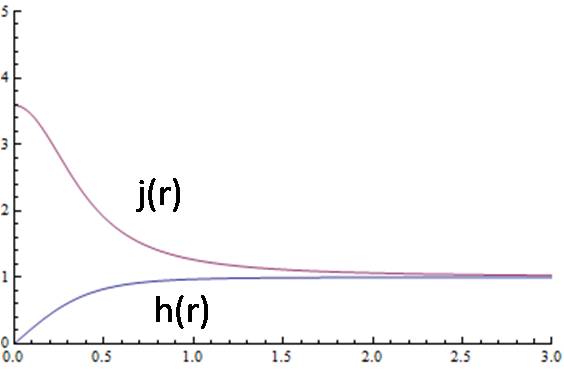}
\end{center}
\caption{\small The Lifshitz star with the smallest allowed value of $j(0)$, the charge density at 
the origin. The figure on the left shows the metric functions $f(r)$ and $g(r)$, while the figure on 
the right shows the electric field $f(r)$ and the charge density $j(r)$.}
\label{lstar}
\end{figure}

The metric and gauge fields of the Lifshitz star with the lowest magic value of $j_0$ are 
shown in Figure~\ref{lstar}.

\section{Black hole thermodynamics}

In the original AdS/CFT correspondence, finite temperature is studied by replacing 
the AdS$_5$ part of the ten-dimensional spacetime by a five-dimensional 
AdS-Schwarzschild black hole \cite{Witten:1998zw} and we expect black holes to
play a similar role here. 

\begin{figure}
\begin{center}
\includegraphics[
width=4.0in
]
{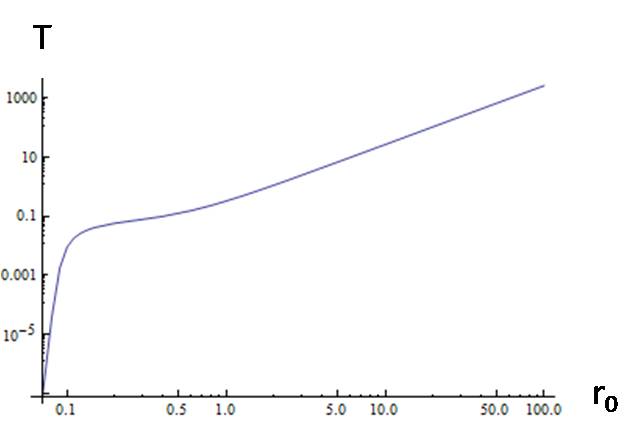}
\end{center}
\caption{\small Numerical results for Hawking temperature as a function of 
black hole size.}
\label{tempfig}
\end{figure}

The Hawking temperature can be obtained by going to a Euclidean metric and 
requiring regularity at the horizon, 
\be
T_H = \frac{1}{4\pi}\frac{f_{0}}{g_{0}}r_{0}^{3}.
\ee
where we have used the near-horizon expansion (\ref{expansion}) for the metric.
The coefficients $g_0$ and $f_0$ are easily determined from our numerical
solutions for the metric and Figure~\ref{tempfig} shows the Hawking temperature
as a function of black hole size. 

We can immediately see important differences between the thermodynamic 
behavior of asymptotically Lifshitz black holes and that of asymptotically AdS 
black holes. Let us first consider black holes that are small compared to the
length scale $L$, set by the cosmological constant. Small AdS black holes 
behave much like ordinary Schwarzschild black holes in asymptotically flat
spacetime. In particular, their Hawking temperature increases as they get
smaller leading to a thermodynamic instability. In the asymptotically Lifshitz 
case, on the other hand, the temperature is a monotonic function of black hole 
size and is rapidly falling at the smallest black hole sizes that our numerical
calculations can handle. This supports our earlier claim that these black 
holes become extremal in the limit of vanishing black hole size. It also means
that there is no analog of the Hawking-Page transition in the Lifshitz system.

\begin{figure}
\begin{center}
\includegraphics[
width=4.0in
]
{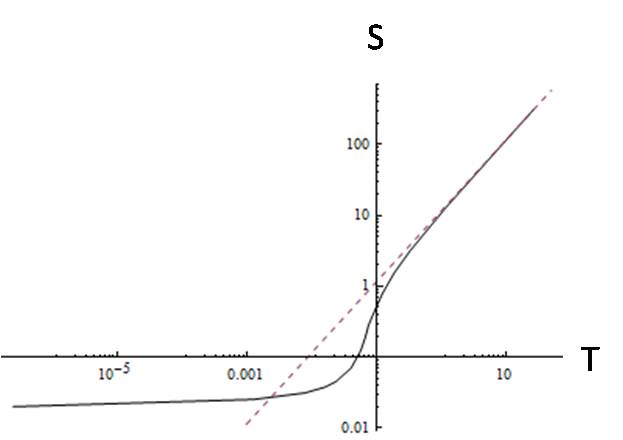}
\end{center}
\caption{\small Black hole entropy as a function of Hawking temperature.}
\label{entropyplot}
\end{figure}

Black holes that are large compared to $L$ satisfy simple scaling relations. 
The form of the near-horizon expansion (\ref{expansion}) suggests that 
$g_0 \sim r_0^{1/2}$ and $f_0\sim r_0^{-1/2}$ for black holes with
$r_0\gg 1$, which in turn gives
\be
S \propto T_H,
\ee
where $S\equiv\pi r_0^2$ is the Bekenstein-Hawking entropy (the corresponding 
behavior for large 3+1 dimensional AdS black holes is $S\propto T_H^2$). 
This is confirmed by our numerical results. We find that
$g_0\approx 0.57\ r_0^{1/2}$ and $f_0\approx 1.96\ r_0^{-1/2}$, giving 
\be
S\approx 11.4\ T_H
\ee
for large asymptotically Lifshitz black holes. Figure~\ref{entropyplot} plots  
the black hole entropy as a function of temperature over the range of
black hole sizes for which we have obtained numerical solutions.

\section{Wilson loops}

The bulk theory we are studying is conjectured to be dual to a boundary theory
at the Lifshitz point. Following \cite{vishwanath-2003} we observe that the term
$K\left(  \nabla^{2}\phi\right)  ^{2}$ in the Lifshitz action (\ref{laction}) can 
be written $K\left\vert \bar{\nabla}\times\bar{E}\right\vert ^{2}$, where 
$E_{i}=\varepsilon_{ij}\partial_{j}\phi$ automatically solves Gauss' law. 
Note also that the field $\phi$ is dimensionless under Lifshitz scaling (\ref{lscaling}),
$K$ is a dimensionless coupling constant, and an additional term in the action of 
the form $\left\vert \bar{E}\right\vert ^{2}$ would be accompanied by a coupling
with dimensions of mass squared. We conclude that the boundary theory can be
viewed as a gauge theory in $2+1$ dimensions with a dimensionless coupling 
constant and an unusual action. The dimensionless coupling suggests 
that the theory perhaps has some nice features in common with conventional 
gauge theory in $3+1$ dimensions. 

With the dual gauge theory in mind, we introduce Wilson loops on the gravity side.
The Wilson loops contain information about the force acting between `quarks',
{\it i.e.} particles that are charged under the gauge fields. The recipe 
given in \cite{Maldacena:1998im,Rey:1998ik} involves hanging a string from 
the boundary, with the end points of the string representing the quarks.

The action of the string for a rectangular Wilson loop, with initial and final
Euclidean time separated by $\Delta$, is given by
\begin{eqnarray}
S  &=& \frac{1}{2\pi\alpha^{\prime}}\int d\sigma d\tau\sqrt{\det
g_{MN}\partial_{\alpha}X^{M}\partial_{\beta}X^{N}} \nonumber \\
&=&  \frac{1}{2\pi\alpha^{\prime}}\int dxdt
\sqrt{g_{tt}\left(  g_{xx}+g_{rr}\left( \frac{dr}{dx}\right)^2\right)}  \nonumber \\
&=&  \frac{L^2 \Delta }{2\pi\alpha^{\prime}} \int dx
\sqrt{f^{2}g^{2}r^{2z-2}\left( \frac{dr}{dx}\right)^{2}+f^{2}r^{2z+2}},
\end{eqnarray}
where we have put $\sigma=x$ \ and $\tau=t$ in static gauge. Extremizing the action 
leads to
\be
\frac{f^{2}r^{2z+2}}{\sqrt{f^{2}g^{2}r^{2z-2}\left( \frac{dr}{dx}\right)^{2}+f^{2}r^{2z+2}
}}=f_{\min}r_{\min}^{z+1},
\ee
where $r_{\min}$ is the $r$ coordinate of the midpoint of the hanging string and
$f_{\min}\equiv f(r_{\min})$. 
The boundary distance between the end points
of the string is then given by
\be
\ell=2\int_{r_{\min}}^{\infty}\frac{dr}{r^{2}}\frac{g}{\sqrt{\left(  
\frac{r}{r_{\min}}\right)^{2z+2}\left(\frac{f}{f_{\min}}\right)^2-1}}.
\ee
The energy of the string configuration is 
\be
V=\frac{L}{2\pi\alpha^\prime}\left[ 
2\int_{r_{\min}}^{\infty}dr \frac{r^{z-1}f g}{
\sqrt{1- \left( \frac{r_{\min}}{r}\right)^{2z+2}\left(\frac{f_{\min}}{f}\right)^2}}
-2\int_{r_{0}}^{\infty}dr\,r^{z-1}f g\right], 
\label{pottemp}
\ee
where we have regularized the expression by subtracting the contribution of two 
straight strings going from the boundary down to the horizon of the black hole 
at $r=r_0$. 

\begin{figure}
\begin{center}
\includegraphics[
width=2.9in
]
{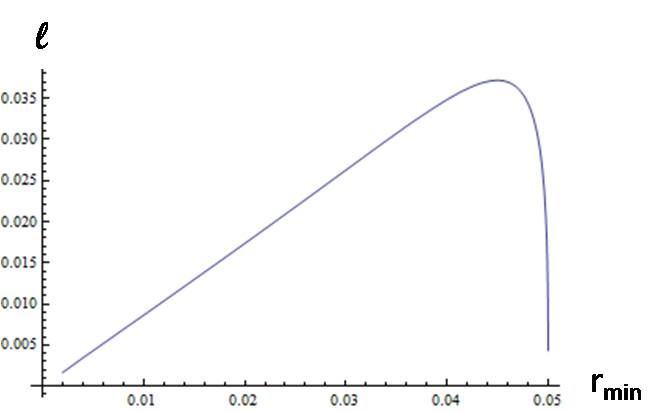}
\includegraphics[
width=2.9in
]
{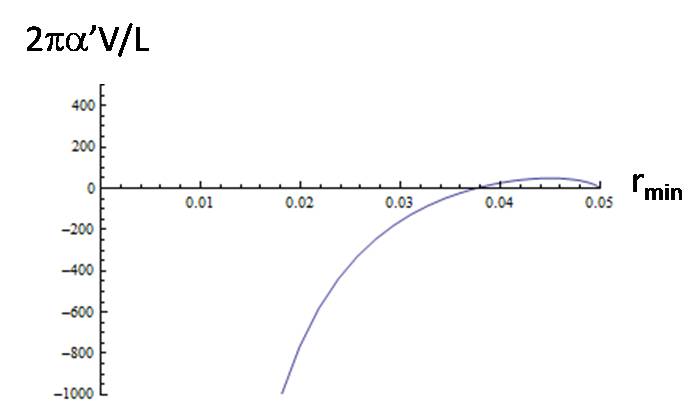}
\end{center}
\caption{\small Screening behavior at finite temperature. The figure on the left shows the 
boundary distance between the endpoints of a hanging string as a function of $r_{\min}$, 
while the figure on the right plots the potential energy as a function of $r_{\min}$.}
\label{landV}
\end{figure}

At vanishing temperature there is no black hole and the above expressions 
simplify considerably.  In this case $f,g\rightarrow1$ and $r_0\rightarrow 0$, and
the boundary distance between the endpoints of the strings is given by
\be
\ell=2\int_{r_{\min}}^{\infty}\frac{dr}{r^{2}}\frac{1}{\sqrt{\left(  
\frac{r}{r_{\min}}\right)  ^{2z+2}-1}}=\frac{2}{r_{\min}}\int_{1}^{\infty}\frac{dy}
{y^{2}}\frac{1}{\sqrt{y^{2z+2}-1}}=\frac{2\sqrt{\pi}}{r_{\min}}
\frac{\Gamma\left(  \frac{2+z}{2+2z}\right)  }{\Gamma\left(  \frac{1}{2+2z}\right)
},
\label{rminell}
\ee
while the potential energy reduces to
\begin{eqnarray}
V  &=&  \frac{L r_{\min}^{z}}{\pi\alpha^{\prime}}\left[  \int_{1}^{\infty}
dy \,y^{z-1}\left(\frac{1}{\sqrt{1-y^{-2z-2}}}-1\right)  -\frac{1}{z}\right] \nonumber \\
&=& -\frac{L^{2}}{\alpha^{\prime}}\frac{1}{L}
\left(\frac{2\sqrt{\pi}}{\ell}\right)^{z}\frac{1}{\sqrt{\pi}z}
\left(  \frac{\Gamma\left(  \frac{2+z}{2+2z}\right)
}{\Gamma\left(  \frac{1}{2+2z}\right)  }\right)^{z+1}.
\label{ellV}
\end{eqnarray}
Here 
$\frac{L^{2}}{\alpha^{\prime}}$ is a dimensionless coupling constant that we
assume to be large. The energy $V$ has dimensions of inverse length, hence the 
factor of $1/L$, and the dependence on $\ell$ can be traced to the 
unconventional scaling properties of the Lifshitz system. At $z=1$ we recover the
results of \cite{Maldacena:1998im,Rey:1998ik}. At $z=2$ it is tempting to relate 
the dimensionless coupling to the Lifshitz coupling $K$. One has to keep in mind, 
though, that our calculation can at best be expected to make sense at strong coupling
whereas $K$ is defined in a free theory.

\begin{figure}
\begin{center}
\includegraphics[
width=2.9in
]
{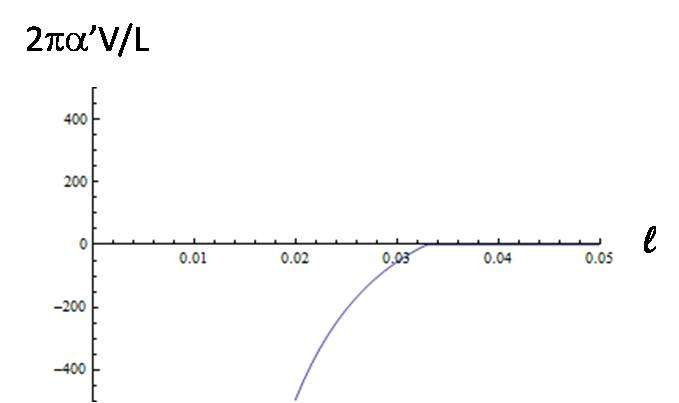}
\includegraphics[
width=2.9in
]
{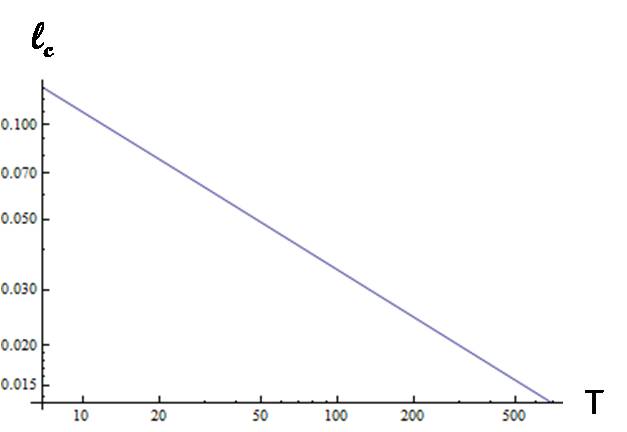}
\end{center}
\caption{\small The figure on the left shows the potential energy as a function of 
the boundary distance between quarks represented by the endpoints of the hanging
string. The figure on the right plots the critical distance, $\ell_c$, where screening 
sets in, {\it vs.} temperature. The log-log graph shows a clear $T^{-1/2}$ dependence.}
\label{Vvsell}
\end{figure}

At non-zero temperature we have to resort to a numerical evaluation of the
integrals, using our numerically evalutated metric as input. Figure~\ref{landV} shows
the boundary distance between the string endpoints as a function of $r_{\min}$ 
and the potential energy of the hanging string, also as function of $r_{\min}$, for
$z=2$. The black hole is taken to be large so that $\ell\ll r_0$.
The figure shows qualitatively the same behavior as was found for $3+1$ dimensional 
gauge theory in \cite{Brandhuber:1998bs,Rey:1998bq}. 
At small separation the inter-quark potential is similar to the zero temperature 
potential but when the separation between the quarks becomes sufficiently large the 
gauge interaction is screened and the potential energy vanishes. The crossover
occurs where the potential energy $V$ becomes positive in Figure~\ref{landV}, 
signaling that at large separation the configuration minimizing the energy is simply 
two straight strings stretching from the boundary down to the horizon. Since there 
is no interaction between such strings the potential simply vanishes in this case.
Figure~\ref{Vvsell} plots the resulting potential as a function of the boundary 
distance between the string endpoints, and also shows how the crossover distance, 
where screening sets in, depends on temperature. The Lifshitz scaling is apparent
in the $\ell_c \propto T^{-1/2}$ falloff as opposed to the $T^{-1}$ behavior seen in
3+1 dimensional gauge theory.

We do not know whether the finite temperature behavior exhibited by the Wilson
loops is in some way related to the ultra-locality discussed in 
\cite{Kachru:2008yh,Ghaemi-2005}. We hope to return to this issue 
in future work.

\section{Conclusions}

In this paper we have explored further the recently proposed gravity dual 
description of Lifshitz type fixed points. We have mainly focused on the 
gravity side of the duality, finding non-trivial spacetime geometries that are 
asymptotic to the Lifshitz fixed point geometry, including black holes that
provide a dual description of a Lifshitz system at finite temperature. 

It is by no means obvious how to incorporate the bulk metrics that we have found 
into a solution of ten dimensional string theory. It is nevertheless tempting to 
proceed under the assumption that such a construction can be found, and that 
a duality analogous to AdS/CFT actually exists. Alternatively, we can view the
gravity dual as a purely phenomenological description of the $2+1$ dimensional
physics. Either way, one is motivated to study the gravitational theory in more
detail.

Our numerical black hole solutions in principle contain all the information
that is needed to calculate finite temperature correlation functions in the
dual system but this requires rather delicate numerical analysis which we
leave for future work. It also requires a better understanding of holographic
renormalization for non-Lorentz invariant field theories \cite{Taylor:2008tg}.

It would also be interesting to make contact with recent work 
\cite{Horava:2008jf} on non-relativistic, non-abelian gauge theories which
exhibit $z=2$ quantum critical behavior. A gravity dual of a large-N limit of 
such a theory would be on firmer theoretical ground than the models 
studied here.

\section*{Acknowledgments}

The work was supported in part by the Swedish Research Council (VR),  
the Icelandic Research Fund, and the University of Iceland Research Fund.

\end{document}